\documentclass[aps,twocolumn,showpacs,preprintnumbers,amsmath,amssymb,superscriptaddress,floatfix,nofootinbib]{revtex4}

\usepackage{graphicx}
\usepackage{epsfig}
\usepackage{epstopdf}
\usepackage{hyperref}
\usepackage{amsmath}
\allowdisplaybreaks[1]
\usepackage{amsfonts}
\usepackage{amssymb}
\usepackage{multirow}
\usepackage{makecell}
\usepackage[dvipsnames]{xcolor}
\usepackage{ulem}
\usepackage{longtable}
\usepackage{array}

\newcommand{\be}{\begin{equation}}
\newcommand{\ee}{\end{equation}}
\newcommand{\bea}{\begin{eqnarray}}
\newcommand{\eea}{\end{eqnarray}}
\newcommand{\bean}{\begin{eqnarray*}}
\newcommand{\eean}{\end{eqnarray*}}

\begin{document}

\title{Decay behaviors of possible $\Lambda_{c\bar{c}}$ states in hadronic molecule pictures}

\author{Chao-Wei Shen} \email{shencw@ucas.ac.cn}
\affiliation{University of Chinese Academy of Sciences (UCAS), Beijing 100049, China}

\author{Jia-Jun Wu} \email{wujiajun@ucas.ac.cn}
\affiliation{University of Chinese Academy of Sciences (UCAS), Beijing 100049, China}

\author{Bing-Song Zou} \email{zoubs@itp.ac.cn}
\affiliation{University of Chinese Academy of Sciences (UCAS), Beijing 100049, China}
\affiliation{Key Laboratory of Theoretical Physics, Institute of Theoretical Physics,
Chinese Academy of Sciences, Beijing 100190,China}
\affiliation{School of Physics, Central South University, Changsha 410083, China}
\date{\today}

\begin{abstract}
In 2010, $\Lambda^*_{c\bar{c}}$ states were predicted as the strange number $S=-1$ partners of $N^*_{c\bar{c}}$, which are well known now as the $P_c$ states and observed experimentally by LHCb Collaboration.
We analyze the decay behaviors of $\Lambda_{c\bar{c}}$  as S-wave hadronic molecules within the effective Lagrangian framework by a similar method, which has been applied on $P_c$ states successfully.
With partial widths of possible decay channels calculated, we find that $\Lambda_{c\bar{c}}(4213)$ and $\Lambda_{c\bar{c}}(4403)$, which are formed as pseudoscalar meson baryon molecules, mainly decay to the $\eta_c \Lambda$ channel.
For the two vector meson baryon molecule states, our results show that the total decay width with $J^P=\frac12^-$ is by one order of magnitude larger than that with $J^P=\frac32^-$.
The decay patterns and relative decay ratios are very different for $\Lambda_{c\bar{c}}(4370)$ being a $D_s^{* -} \Lambda_c^+$ or $\bar{D}^{*} \Xi_c$ molecule state.
The main decay channels of $\Lambda_{c\bar{c}}(4550)$ are $\bar{D}^{(*)} \Xi^{(*,\prime)}_c$ because of the pseudoscalar meson exchange mechanism.
In addition, $\bar{D}^{*} \Xi_c$ is the dominant decay channel of $\Lambda_{c\bar{c}}(4490)$ which is assumed as a $\bar{D} \Xi_c^{*}$ bound state.
These decay patterns of the $\Lambda^*_{c\bar{c}}$ states would provide a guidance for their future experimental searches and  help us to understand their internal structures.

\end{abstract}

\maketitle

\section{Introduction}


In Ref.~\cite{Wu:2010jy},
within the hidden local symmetry, not only $N_{c\bar{c}}$ states but also $\Lambda_{c\bar{c}}$ states are predicted.
The predicted $N_{c \bar{c}}$ states are $\bar{D}\Sigma_c$ or $\bar{D}^* \Sigma_c$ S-wave bound states which locate around 4.3~GeV.
They are found to be consistent with the observed three peak structures by LHCb Collaboration in 2019~\cite{Aaij:2019vzc}.
In the LHCb earlier paper~\cite{Aaij:2015tga}, they named such states as $P_c$, whose flavor quanta number is the same as $N^*$ but definitely have $c\bar{c}$ components.
However, the newest results with higher statistic data from LHCb group have not been partial wave analysed, thus the spin and parity of these states are still unknown.
The $P_c$ states have attracted much attention since the first proposal in 2010~\cite{Wu:2010jy}, and become a very hot topic once the signals of them were first seen in 2015 by LHCb~\cite{Aaij:2015tga}.
The main reason is that they are the first exotic baryons discovered experimentally.
But until now the only experimental information of these states comes from $J/\psi p$ invariant spectrum.
Only their masses and total widths can be extracted.
The quantum numbers and the internal structures of these states are still unknown.
Many models have been applied to study them and various explanations are proposed~\cite{Chen:2016qju,Guo:2017jvc}.
Roughly speaking, there are three different views of these peaks.
Firstly, they are recognized as meson baryon molecular states, which can be divided in anticharmed meson charmed baryon states~\cite{Wu:2010jy,Yang:2011wz,Wu:2012md, Oset:2012ap,Garcia-Recio:2013gaa,Xiao:2013yca,Uchino:2015uha, Chen:2015loa, Chen:2015moa, Roca:2015dva, He:2015cea, Huang:2015uda, Wang:2015qlf, Yang:2015bmv, Chen:2016heh, Roca:2016tdh, Lu:2016nnt, Shimizu:2016rrd, Shen:2016tzq, Ortega:2016syt,Meissner:2015mza, Yamaguchi:2016ote, He:2016pfa, Oset:2016nvf, Xiao:2016ogq, Lin:2017mtz, Yamaguchi:2017zmn, Shen:2017ayv, Lin:2018kcc, Shimizu:2019jfy}, baryo-charmonium states~\cite{Anwar:2018bpu, Eides:2018lqg}, or the mixture of them~\cite{Burns:2015dwa,Takeuchi:2016ejt}.
Secondly, they are considered in constituent quark model~\cite{Wang:2011rga, Xiang:2017byz, Li:2017kzj, Hiyama:2018ukv}, diquark-diquark-antiquark picture~\cite{Yuan:2012wz, Chen:2016otp, Lebed:2015tna, Li:2015gta,Wang:2015epa}, and diquark-triquark picture~\cite{Zhu:2015bba}.
Thirdly, the narrow peak of $P_c(4450)$ might result from triangle singularity (TS) effect~\cite{Guo:2015umn, Liu:2015fea, Guo:2016bkl}, which is purely kinematic effect, although for some quantum numbers of the $P_c$ state preferred in Ref.~\cite{Aaij:2015tga}, such as $3/2^-$ or $5/2^+$, the TS can not explain the peak as shown in Ref.~\cite{Bayar:2016ftu}.
Recently, the new updated results by LHCb collaboration~\cite{Aaij:2019vzc} clearly show that the three narrow states are all just below the corresponding anticharmed meson charmed baryon thresholds, which strongly suggests the hadronic molecule nature for them.
There are several new theoretical papers~\cite{Chen:2019asm, Chen:2019bip, Liu:2019tjn, Guo:2019fdo, He:2019ify, Liu:2019zoy, Huang:2019jlf, Xiao:2019mvs, Shimizu:2019ptd, Guo:2019kdc} triggered by the new LHCb results support meson baryon state description, mainly from $\bar{D}^{(*)}\Sigma^{(*)}$, although they have different views in detail.
Only Ref.~\cite{Ali:2019npk} argues that the internal structures still rely on the parities of these states.
Here, we want to emphasize that the mass and total width are maybe not enough to distinguish among various models.
Definitely more information, such as the spin and parity, and partial decay widths of these states are needed from the new measurements in experiments.
Correspondingly, it is worthy to make the prediction of the partial widths of these states from the theoretical side to help experimentalists to find new reactions to search these states.
%

In previous papers~\cite{Lu:2016nnt,Shen:2016tzq,Lin:2017mtz,Wang:2015qlf}, the decay patterns of $P_{c}$ based on different assumptions of their internal structures were studied.
It is found that if $P_c^+(4380)$ is a $J^P=\frac32^-$ $\bar{D}^* \Sigma_c$ molecular state, its width will be around 50 MeV, which is much smaller than that analyzed from the experimental data.
However, the width would be around 150 MeV if $P_c^+(4380)$ is assumed as a $J^P=\frac32^-$ $\bar{D} \Sigma^*_c$ molecular state.
This implies that it has more possibilities to be a $\bar{D} \Sigma^*_c$ molecular state with $J^P=\frac32^-$.
On the other hand, $P_c^+(4450)$ is more likely to be a $\bar{D}^* \Sigma_c$ molecule with $J^P=\frac52^+$.
In the updated results from Ref.~\cite{Aaij:2019vzc}, the $P_c^+(4380)$ is not mentioned but this broad structure still exists in the fit, and $P_c^+(4450)$ splits into two states, $P_c^+(4440)$ and $P_c^+(4457)$, with both their spin and parity not yet decided.
The width of $P_c^+(4440)$ is around 20 MeV which is still comparable with the calculations from Refs.~\cite{Lu:2016nnt,Shen:2016tzq,Lin:2017mtz,Wang:2015qlf}.
While another new state $P_c(4312)$ is more likely to be a $\bar{D}\Sigma_c$ molecular state and is missing in the earlier measurement.
Correspondingly, in the previous relevant references, $P_c(4312)$ has not yet been studied.
In the present work, we consider the decay behaviors of the predicted $\Lambda_{c\bar{c}}$ states similarly, which should be quite useful for understanding their nature with the help of the forthcoming experiments.

%
The $\Lambda_{c\bar{c}}$, which are partners of $P_c$ with strangeness -1 states, have already been investigated by several groups.
The strange hidden-charm state is considered in the quark model, and both color octet type and color singlet type are studied~\cite{Irie:2017qai}, which is different from this work, where the $\Lambda_{c\bar{c}}$ is assumed as a molecule with two color singlet parts.
It is found that the $\Lambda_{c\bar{c}}$ below 4400 MeV are all formed as two color octet parts, color octet $uds$ and $c\bar{c}$ parts, and the spin and parity can be $1/2^-$ and $3/2^-$.
Several possible decay modes are then discussed and they found that both $\eta_c \Lambda$ and $J/\psi \Lambda$ channel are suppressed, while $\bar{D}_s\Lambda_c$ and $\bar{D}\Xi_c$ are the possible main decay channels.
Ref.~\cite{Chen:2016ryt} calculated the $\Lambda_c \bar{D}_s^*$, $\Sigma_c^{(*)} \bar{D}_s^*$ and $\Xi_c^{(\prime,*)} \bar{D}^*$ interactions in the one boson exchange model.
The results tell that a $\Xi_c^\prime \bar{D}^*$ state with $J^P=\frac12^-$ and two $\Xi_c^* \bar{D}^*$ states with $J^P=\frac12^-$ and $\frac32^-$ are the most promising molecular states.
In addition, the production of $\Lambda^*_{c\bar{c}}$ are predicted in various decay, and all the results suggest to search it in the $J/\psi \Lambda$ invariant spectrum.
Refs.\cite{Feijoo:2015kts,Oset:2016nvf} studied the $\Lambda_b \to J/\psi \eta \Lambda$ decay,
while Refs.~\cite{Chen:2015sxa, Oset:2016nvf} make the prediction of  $\Xi_b^- \to J/\psi K^- \Lambda$ decay, and in Ref.~\cite{Lu:2016roh,Oset:2016nvf}, a theoretical study of the $\Lambda_b \to J/\psi K^0 \Lambda$ reaction is performed.
Within the theoretical uncertainties, all of these studies show that there would be rather stable signals of the hidden-charm strange states.
And the partial widths are all consistent with the predictions in Ref.\cite{Wu:2010jy}, where only the contribution of vector meson exchange is considered.
%

%
Among the predicted states in Ref.~\cite{Wu:2010jy}, there are six $\Lambda_{c\bar{c}}$ states.
Two of them are from pseudoscalar meson baryon (PB) channel. $\Lambda_{c\bar{c}}(4213)$ is coupled to both $\bar{D}_s \Lambda_c^+$ and $\bar{D} \Xi_c$ channels, while $\Lambda_{c\bar{c}}(4403)$ only couples to $\bar{D} \Xi_c^\prime$ channel.
The other four states are from vector meson baryon (VB) channel, two of which are around 4370~MeV and couple to both $\bar{D}_s^* \Lambda_c^+$ and $\bar{D}^* \Xi_c$ channels, and the other two $\Lambda_{c\bar{c}}$ states only couple to $\bar{D}^* \Xi_c^\prime$ channel, with masses around 4550~MeV.
Note that in Ref.~\cite{Wu:2010jy} for each VB case its bound states always appear as a degenerate pair of spin-parity $J^P=1/2^-$ and $3/2^-$, respectively, due to an approximation of neglecting small spin-dependent force.
We consider $\Lambda_{c\bar{c}}(4213/4370)$ as either a $\bar{D}_s^{(*)} \Lambda_c^+$ or a $\bar{D}^{(*)} \Xi_c$ S-wave bound state and $\Lambda_{c\bar{c}}(4403/4550)$ as a $\bar{D}^{(*)} \Xi_c^\prime$ S-wave molecule.
An additional $\bar{D} \Xi_c^*$ S-wave bound state is also taken into consideration assuming to be $\Lambda_{c\bar{c}}(4490)$ with spin-parity-$3/2^-$ and binding energy of about 23~MeV.
In the work, we make an estimation of the partial decay widths of these seven $\Lambda_{c\bar{c}}$ states to possible two body decay channels, which is expected to help figure out the nature of these $\Lambda_{c\bar{c}}$ states.
%
%
%


This article is organized as follows. In Sect.~\ref{sec:formalism}, we present the theoretical framework of our calculation. In Sect.~\ref{sec:result}, the numerical decay widths of the $\Lambda_{c\bar{c}}$ states and relevant discussions about these results are presented, then a brief summary in Sect.~\ref{sec:summary} of this work is followed and an Appendix.~\ref{AppendixA} is presented at last.

\section{Theoretical framework} \label{sec:formalism}

\begin{figure}[htbp]
\centering
\includegraphics[scale=1.0]{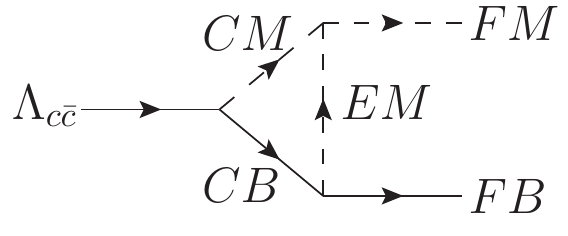}
\caption{Feynman diagram for the two body decay of $\Lambda_{c \bar{c}}$ in the molecule picture, where CM(CB) denotes the constituent meson (baryon) of the composite system, FM(FB) denotes the final meson (baryon), EM denotes the exchanged meson.
\label{Fig:feyndiag}}
\end{figure}

The decays of these $\Lambda_{c\bar{c}}$ states proceed through triangular diagrams as shown in Fig.~\ref{Fig:feyndiag}.
The possible molecular assumptions, their decay modes and corresponding exchanged mesons are listed in Table~\ref{table:mode}.

\begin{table*}
\caption{\label{table:mode}All possible decay modes considered in the calculation. NA denotes that the corresponding decay is not allowed.}
\renewcommand\arraystretch{1.5}
\begin{tabular}{l|c|ccccccc}
\hline
\multirow{3}*{Mode} & \multirow{3}*{\makecell*{Threshold\\($\mathrm{MeV}$)}} & \multicolumn{7}{c}{Exchanged Particle} \\
\cline{3-9}
& & \multicolumn{2}{c}{$\Lambda_{c \bar{c}}(4213)$} & \multicolumn{1}{c}{$\Lambda_{c \bar{c}}(4403)$} & \multicolumn{2}{c}{$\Lambda_{c \bar{c}}(4370)$} & \multicolumn{1}{c}{$\Lambda_{c \bar{c}}(4490)$} & \multicolumn{1}{c}{$\Lambda_{c \bar{c}}(4550)$} \\
\cline{3-9}
& & \multicolumn{1}{c}{$D_s^{-} \Lambda_c^+(4255)$} & \multicolumn{1}{c}{$\bar{D} \Xi_c(4337)$} & \multicolumn{1}{c}{$\bar{D} \Xi_c^\prime(4445)$} & \multicolumn{1}{c}{$D_s^{* -} \Lambda_c^+(4399)$} & \multicolumn{1}{c}{$\bar{D}^{*} \Xi_c(4478)$} & \multicolumn{1}{c}{$\bar{D} \Xi_c^*(4513)$} & \multicolumn{1}{c}{$\bar{D}^{*} \Xi_c^\prime(4587)$} \\
\hline
$J/\psi \Lambda$			& 4213 &	NA	&	NA	&	$D$, $D^*$	&	$D_s$, $D_s^*$	&	$D$, $D^*$	&	$D$, $D^*$	&	$D$, $D^*$	\\
$\eta_c \Lambda$			& 4100 &	$D_s^*$	&	$D^*$	&	$D^*$	&	$D_s$, $D_s^*$	&	$D$, $D^*$	& $D^*$	&	$D$, $D^*$	\\
$\bar D \Xi_c$				& 4337 &	NA	&	NA	&	NA	&	$K^*$	&	$\eta_c$, $\rho$, $\omega$, $J/\psi$	&	NA	&	$\pi$, $\eta$	\\
$D_s^- \Lambda_c^+$		& 4255 &	NA	&	NA	&	NA	&	$\eta_c$, $J/\psi$	&	$\bar K^*$	&	NA	&	$\bar K$		\\
$\phi \Lambda$				& 2135 &	$D_s$, $D_s^*$	&	NA	&	NA	&	$D_s$, $D_s^*$	&	NA	&	NA	&	NA	\\
$\rho \Sigma$				& 1968 &	NA	&	$D$, $D^*$	&	$D$, $D^*$	&	NA	&	$D$, $D^*$	&	$D$, $D^*$	&	$D$, $D^*$	\\
$\omega \Lambda$			& 1898 &	NA	&	$D$, $D^*$	&	$D$, $D^*$	&	NA	&	$D$, $D^*$	&	$D$, $D^*$	&	$D$, $D^*$	\\
$\pi \Sigma$				& 1331 &	NA	&	$D^*$	&	$D^*$	&	NA	&	$D$, $D^*$	&	$D^*$	&	$D$, $D^*$	\\
$\eta \Lambda$				& 1664 &	$D_s^*$	&	$D^*$	&	$D^*$	&	$D_s$, $D_s^*$	&	$D$, $D^*$	& $D^*$	&	$D$, $D^*$	\\
$\eta^\prime \Lambda$		& 2073 &	$D_s^*$	&	$D^*$	&	$D^*$	&	$D_s$, $D_s^*$	&	$D$, $D^*$	&	$D^*$	&	$D$, $D^*$	\\
$\bar K N$				& 1435 &	$D^*$	&	NA	&	NA	&	$D$, $D^*$	&	NA	&	NA	&	NA	\\
$\bar K^* N$				& 1833 &	$D$, $D^*$	&	NA	&	NA	&	$D$, $D^*$	&	NA	&	NA	&	NA	\\
$K \Xi$					& 1814 &	NA	&	$D_s^*$	&	$D_s^*$	&	NA	&	$D_s$, $D_s^*$	&	$D_s^*$	&	$D_s$, $D_s^*$	\\
$K^* \Xi$					& 2212 &	NA	&	$D_s$, $D_s^*$	&	$D_s$, $D_s^*$	&	NA	&	$D_s$, $D_s^*$	&	$D_s$, $D_s^*$	&	$D_s$, $D_s^*$	\\
$D_s^{* -} \Lambda_c^+$		& 4399 &	NA	&	NA	&	$\bar K$	&	NA	&	NA	&	$\bar K$	&	$\bar K$	\\
$\bar D^* \Xi_c$			& 4478 &	NA	&	NA	&	NA	&	NA	&	NA	&	$\pi$, $\eta$	&	$\pi$, $\eta$	\\
$\bar D \Xi_c^*$			& 4513 &	NA	&	NA	&	NA	&	NA	&	NA	&	NA	&	$\pi$, $\eta$, $\eta^\prime$, $\rho$, $\omega$ \\
$\bar D \Xi_c^\prime$		& 4445 &	NA	&	NA	&	NA	&	NA	&	NA	&	$\rho$, $\omega$	&	$\pi$, $\eta$, $\eta^\prime$, $\eta_c$, $\rho$, $\omega$, $J/\psi$	\\
\hline
\end{tabular}
\end{table*}

We use effective Lagrangian method to calculate all the considered processes and the involved Lagrangians of various kinds of vertices are given~\cite{MuellerGroeling:1990cw, Molina:2008jw, Shen:2017ayv}:
\begin{eqnarray}\label{Lag:all}
\mathcal{L}_{BBP} &=& \frac{g_{BBP}}{m_P} \bar B \gamma^\mu \gamma^5 \partial_\mu P B, \nonumber \\
\mathcal{L}_{BBV} &=& - g_{BBV} \bar B \gamma^\mu V_\mu B, \nonumber \\
\mathcal{L}_{BDP} &=& \frac{g_{BDP}}{m_P} (\bar D^\mu B + \bar B D^\mu)\partial_\mu P, \nonumber \\
\mathcal{L}_{BDV} &=& -i \frac{g_{BDV}}{m_V} (\bar D^\mu \gamma^5 \gamma^\nu B - \bar B \gamma^5 \gamma^\nu D^\mu) \nonumber \\
&& \times (\partial_\mu V_\nu - \partial_\nu V_\mu), \nonumber \\
\mathcal{L}_{PPV} &=& - g_{PPV}(P \partial_\mu P - \partial_\mu P P)V^\mu, \nonumber \\
\mathcal{L}_{VVP} &=& \frac{g_{VVP}}{m_V} \epsilon_{\mu \nu \alpha \beta} \partial^\mu V^\nu \partial^\alpha V^\beta P, \nonumber \\
\mathcal{L}_{VVV} &=& g_{VVV} < (\partial_\mu V_\nu -\partial_\nu V_\mu)V^\mu V^\nu >,
\end{eqnarray}
where $P$, $V$, $B$, $D$ denote pseudoscalar, vector meson, octet and decuplet baryon, respectively.
It should be mentioned that we use the masses of pseudoscalar and vector mesons in the Lagrangians instead of $m_\pi$, $m_\rho$ and $m_\omega$ in the original expressions.
This could be regarded as a correction to the strongly broken SU(4) flavor symmetry, which is applied in calculating the coupling constants.
Then we apply SU(4) flavor symmetry and hidden local symmetry to relate the coupling constants in Eq.~(\ref{Lag:all}) to known couplings.
As shown in Ref.~\cite{Wu:2010jy, Hofmann:2005sw}, hidden local symmetry will make $\bar{D}\Xi_c$ and $\bar{D}\Xi_c^\prime$ decouple, since $\Xi_c$ and $\Xi_c^\prime$ belong to $\bf 6$ and $\bf \bar{3}$ of $qq$ (two light quarks pair) components, respectively.
The relation of all the needed couplings constants and the values of given couplings are shown in Appendix~\ref{AppendixA}.
The S-wave interactions involving $\Lambda_{c \bar{c}}$ is taken into consideration by using Lorentz covariant orbital-spin (L-S) scheme~\cite{Zou:2002yy}, and the corresponding Lagrangians from the L-S scheme are:
\begin{eqnarray} \label{Lag:first}
\mathcal{L}_{\Lambda_{c \bar{c}}(\frac{1}{2}^-)PB} &=& g_{\Lambda_{c \bar{c}}(\frac{1}{2}^-)PB} \bar{\Lambda}_{c \bar{c}} P B, \nonumber \\
\mathcal{L}_{\Lambda_{c \bar{c}}(\frac{1}{2}^-)VB} &=& g_{\Lambda_{c \bar{c}}(\frac{1}{2}^-)VB} \bar{\Lambda}_{c \bar{c}} \gamma_5 (g_{\mu\nu}-\frac{p_\mu p_\nu}{p^2}) \gamma^\nu V^\mu B, \nonumber \\
\mathcal{L}_{\Lambda_{c \bar{c}}(\frac{3}{2}^-)VB} &=& g_{\Lambda_{c \bar{c}}(\frac{3}{2}^-)VB} \bar{\Lambda}_{c \bar{c} \mu} V^\mu B, \nonumber \\
\mathcal{L}_{\Lambda_{c \bar{c}}(\frac{3}{2}^-)PD} &=& g_{\Lambda_{c \bar{c}}(\frac{3}{2}^-)PD} \bar{\Lambda}_{c \bar{c} \mu} P D^\mu,
\end{eqnarray}
where $p_\mu$ is the momentum of the $\Lambda_{c \bar{c}}$ state.

The coupling constants in Eq.~(\ref{Lag:first}) can be estimated by using~\cite{Weinberg:1965zz, Baru:2003qq, Guo:2008zg}:
\begin{equation}\label{eq:g1}
g^2=\frac{4\pi}{4Mm_2}\frac{(m_1+m_2)^{5/2}}{(m_1m_2)^{1/2}}\sqrt{32\epsilon}
\end{equation}
where $M$, $m_1$ and $m_2$ are the masses of $\Lambda_{c \bar{c}}$, constituent meson and constituent baryon, respectively, and $\epsilon$ is the binding energy.
It should be noticed that Eq.~(\ref{eq:g1}) is valid for an S-wave shallow bound state.
%
%
In Table~\ref{table:Lambdacccoupling}, all these $\Lambda^*_{c\bar{c}}$ involved coupling constants are listed, and their corresponding values in Ref.~\cite{Wu:2010jy} are also given.
It shows that the coupling constants determined in these two methods are quite close, except the values of those coupled to $D_s^{(*)\,-}\Lambda_c^+$ channel.
%
The difference is that there are coupled channel effects in Ref.~\cite{Wu:2010jy}, while these coupling constants in the present work is calculated for a specific bound state.

\begin{table}
\caption{\label{table:Lambdacccoupling}{All the $\Lambda^*_{c\bar{c}}$ involved coupling constants from Eq.(\ref{eq:g1})/Ref.~\cite{Wu:2010jy}.} }
\renewcommand\arraystretch{1.5}
\begin{tabular}{cccc}
\hline
{$\Lambda^*_{c\bar{c}}$}
& \multicolumn{3}{c}{Consistent Channel} \\
\hline
    & \multicolumn{1}{c}{$D_s^{-} \Lambda_c^+$} & \multicolumn{1}{c}{$\bar{D} \Xi_c$}
  & \multicolumn{1}{c}{$\bar{D} \Xi_c^\prime$}  \\
\ \ $\Lambda^*_{c\bar{c}}(4213)$ \ \	& \ 2.58/1.37 \ &	\ 3.32/3.25 \ &	-	 	        \\
$\Lambda^*_{c\bar{c}}(4403)$  & -        &	-        &	\ 2.50/2.64 \	 	\\
\hline
  & \multicolumn{1}{c}{$D_s^{*\,-} \Lambda_c^+$} & \multicolumn{1}{c}{$\bar{D}^* \Xi_c$}
  & \multicolumn{1}{c}{$\bar{D}^* \Xi_c^\prime$}  \\
$\Lambda^*_{c\bar{c}}(4370)$	& 2.36/1.23 &	3.21/3.14 &	-	 	        \\
$\Lambda^*_{c\bar{c}}(4550)$  & -        &	-        &	2.40/2.53	 	\\
\hline
  & \multicolumn{1}{c}{$D_s^{-} \Xi_c^*$}  \\
$\Lambda^*_{c\bar{c}}(4490)$	& 2.10 	\\
\hline
\end{tabular}
\end{table}

Since there exists ultraviolet (UV) divergences in the loop integrals when calculating the amplitudes, we use the same method as Refs.~\cite{Shen:2016tzq,Lin:2017mtz,Lin:2018kcc} to absorb the divergence.
A Gaussian regulator with the cutoff $\Lambda_1$ is used to suppress the contribution of the two constituents at short distance and another off-shell form factor is included for the exchanged meson with the cutoff $\Lambda_2$.
Their explicit forms are given as:
\begin{eqnarray}
\label{eq:FF}
\phi(q_E^2/\Lambda_1^2) &=& {\rm{exp}}(-q_E^2/\Lambda_1^2), \nonumber \\
f(q^2) &=& \frac{\Lambda_2^4}{(m^2-q^2)^2 + \Lambda_2^4},
\end{eqnarray}
where $q_E=(m_{CM}p_{CB}-m_{CB}p_{CM})/(m_{CM}+m_{CB})$ is the Euclidean Jacobi momentum.
As discussed in our previous work, the cutoff $\Lambda_1$ varies from 0.8 to 1.2~GeV and $\Lambda_2$ is in the range of 1.8 to 2.2~GeV.

\section{Results and discussions} \label{sec:result}

%
%
Taking all into account, the partial decay widths of the seven $\Lambda_{c\bar{c}}$ states in different S-wave hadronic molecular assumptions to possible two body channels listed in Table~\ref{table:mode} could be calculated.
According to the analysis in previous works~\cite{Lin:2017mtz,Lin:2018kcc}, we adopt $\Lambda_1=1.0\ \mathrm{GeV}$ and $\Lambda_2=2.0\ \mathrm{GeV}$ as a set of typical values.
The numerical results obtained with this set of typical cutoff values are displayed in Table~\ref{table:width}.
These values cannot be regarded as the precise results because our model does not include the coupled channel effects and also suffers from large uncertainties due to the coupling constants from SU(4) relations and the choice of cutoffs $\Lambda_1$ and $\Lambda_2$.
The uncertainties of coupling constants would come from the SU(4) breaking, hidden local symmetry breaking and multi-loop contributions, and all of them are beyond the present model and should be improved in the future.
Since the cutoff dependence of the decay widths will not change the relative value of partial widths as shown later, we can pick up the main decay channels from Table~\ref{table:width} to estimate the cutoff dependence.
In Tables~\ref{table:DsLc12}$-$\ref{table:DXcstBE}, we show all partial decay widths with different values of cutoffs $\Lambda_1$ and $\Lambda_2$ varying from $0.8-1.2$~GeV and $1.8-2.2$~GeV, respectively.
We find that the partial widths are rather stable for different choices of $\Lambda_2$, while they will suffer uncertainties of a factor 4 for $\Lambda_1$ from 0.8 to 1.2~GeV.
%
%
It is confirmed that the cutoffs only affect the total decay widths but will not influence the relative decay ratios.

\begin{table*}
\caption{\label{table:width}The partial decay widths of the three $\Lambda_{c\bar{c}}$ states in different S-wave hadronic molecular assumptions to possible decay channels with $\Lambda_1=1.0\ \mathrm{GeV}$ and $\Lambda_2=2.0\ \mathrm{GeV}$. All of the decay widths are in the unit of MeV, and NA denotes that the corresponding decay is not allowed.}
\renewcommand\arraystretch{1.5}
\begin{tabular}{l|*{10}{c}}
\hline
\multirow{4}*{Mode} & \multicolumn{10}{c}{Widths ($\mathrm{MeV}$)} \\
\cline{2-11}
& \multicolumn{2}{c}{$\Lambda_{c \bar{c}}(4213)$} & \multicolumn{1}{c}{$\Lambda_{c \bar{c}}(4403)$} & \multicolumn{4}{c}{$\Lambda_{c \bar{c}}(4370)$} & \multicolumn{1}{c}{$\Lambda_{c \bar{c}}(4490)$} & \multicolumn{2}{c}{$\Lambda_{c \bar{c}}(4550)$} \\
\cline{2-11}
& \multicolumn{1}{c}{$D_s^{-} \Lambda_c^+$} & \multicolumn{1}{c}{$\bar{D} \Xi_c$} & \multicolumn{1}{c}{$\bar{D} \Xi_c^\prime$} & \multicolumn{2}{c}{$D_s^{* -} \Lambda_c^+$} & \multicolumn{2}{c}{$\bar{D}^{*} \Xi_c$} & \multicolumn{1}{c}{$\bar{D} \Xi_c^*$} & \multicolumn{2}{c}{$\bar{D}^{*} \Xi_c^\prime$} \\
\cline{2-11}
& \multicolumn{1}{c}{${\frac12}^-$} & \multicolumn{1}{c}{${\frac12}^-$} & \multicolumn{1}{c}{${\frac12}^-$} & \multicolumn{1}{c}{${\frac12}^-$} & \multicolumn{1}{c}{${\frac32}^-$} & \multicolumn{1}{c}{${\frac12}^-$} & \multicolumn{1}{c}{${\frac32}^-$} & \multicolumn{1}{c}{${\frac32}^-$} & \multicolumn{1}{c}{${\frac12}^-$} & \multicolumn{1}{c}{${\frac32}^-$} \\
\hline
$J/\psi \Lambda$			&	NA	&	NA	&	0.045	&	3.006	&	0.766	&	2.380	&	0.592	&  0.464 &	10.143	&	2.254	\\
$\eta_c \Lambda$			&  2.624 & 1.961 & 9.126 &	0.085	&	0.002	&	0.076	&	0.002	&  0.115 &	0.485	&	0.019	\\
$\bar D \Xi_c$				&	NA	&	NA	&	NA	&	1.679	&	0.002	&	3.260	&	0.002	& NA  & 99.134	&	5.094	\\
$D_s^- \Lambda_c^+$		&	NA	&	NA	&	NA	&	0.000	&	0.000	&	3.862	&	0.021	&  NA  &	4.307	&	0.316	\\
$\phi \Lambda$				&	0.269	&	NA	&	NA	&	4.403	&	0.380	&	NA	&	NA	& NA  &	NA	&	NA	\\
$\rho \Sigma$				&	NA	&	1.245	&	0.248	&	NA	&	NA	&	20.697	&	1.781	&  2.837  &	6.247	&	0.428	\\
$\omega \Lambda$			&	NA	&	0.137	&	0.243	&	NA	&	NA	&	2.304	&	0.194	&  2.824  &	6.248	&	0.417	\\
$\pi \Sigma$				&	NA	&	2.902	&	1.008	&	NA	&	NA	&	0.881	&	0.074	& 0.165 &	0.314	&	0.026	\\
$\eta \Lambda$				&  0.749 & 0.114 & 0.354 &	0.207	&	0.017	&	0.033	&	0.003	& 0.056  &	0.107	&	0.009	\\
$\eta^\prime \Lambda$		&  0.438 & 0.278 & 0.845 &	0.100	&	0.008	&	0.063	&	0.005	&  0.109 &	0.212	&	0.017	\\
$\bar K N$				&	4.417	&	NA	&	NA	&	1.359	&	0.112	&	NA	&	NA	& NA  &	NA	&	NA	\\
$\bar K^* N$				&	1.301	&	NA	&	NA	&	21.103	&	1.537	&	NA	&	NA	& NA  &	NA	&	NA	\\
$K \Xi$					&	NA	&	1.528	&	0.531	&	NA	&	NA	&	0.410	&	0.035	& 0.075 &	0.149	&	0.012	\\
$K^* \Xi$					&	NA	&	0.501	&	0.109	&	NA	&	NA	&	8.463	&	0.824	& 1.130 &	2.575	&	0.204	\\
$D_s^{* -} \Lambda_c^+$		&	NA	&	NA	&	0.112	&	NA	&	NA	&	NA	&	NA	&  1.304 &	2.558	&	0.327	\\
$\bar D^* \Xi_c$			&	NA	&	NA	&	NA	&	NA	&	NA	&	NA	&	NA	&  15.709  & 47.178	&	4.687	\\
$\bar D \Xi_c^*$			&	NA	&	NA	&	NA	&	NA	&	NA	&	NA	&	NA	&  NA  &	1.127	&	24.421	\\
$\bar D \Xi_c^\prime$		&	NA	&	NA	&	NA	&	NA	&	NA	&	NA	&	NA	&  0.024  &	106.219	&	2.028	\\
\hline
Total						&  \ 9.798 \ & \ 8.666 \ & \ 12.621 \ &	\ 31.942 \	&	\ 2.824 \	&	\ 42.429 \	&  \ 3.533 \	& \ 24.812 \  &	\ 287.003 \	&	\ 40.259 \	\\
\hline
\end{tabular}
\end{table*}

In the first two columns of Table~\ref{table:width}, we find that for the $\Lambda_{c \bar{c}}(4213)$ states with spin-parity-$1/2^-$, the total decay widths are about 10 MeV for both $D_s^{-} \Lambda_c^+$ and $\bar{D} \Xi_c$ molecular assumptions, while the decay patterns are very different in these two cases.
For being a $D_s^{-} \Lambda_c^+$ bound state, the three main decay channels are $\bar{K} N$, $\eta_c \Lambda$ and $\bar{K}^* N$, whose ratio reaches 85\%.
However, if $\Lambda_{c \bar{c}}(4213)$ is a $\bar{D} \Xi_c$ molecule, it mostly decays to $\pi \Sigma$, $\eta_c \Lambda$, $K \Xi$ and $\rho \Sigma$.
These four final states account for 88\% of its width.
We consider the dependence of the partial decay widths of these main channels on the cutoffs $\Lambda_1$ and $\Lambda_2$, and the corresponding results are shown in Tables~\ref{table:DsLc12} and \ref{table:DXc12}.
%
%
Furthermore, in the present work we also include the pseudoscalar meson exchange, and we found that the vector meson baryon channels contribute around $20\%$ of the total width.

%
The numerical decay widths are very different from those of Ref.~\cite{Wu:2010jy}.
%
As discussed before, the two main differences between these two works are that in this work the coupled channel effects are not included and the contributions from pseudoscalar meson exchange are lacked in Ref.~\cite{Wu:2010jy}.
For the $\Lambda_{c\bar{c}}(4213)$ case, there is no pseudoscalar meson exchange for pseudoscalar meson baryon decay channels, so it is the nice place to inspect the coupled channel effects.
In Ref.~\cite{Wu:2010jy}, $\Lambda_{c\bar{c}}(4213)$ is a two couple channels bound state, while in this work we treat it as a $D_s^{-} \Lambda_c^+$ and a $\bar{D} \Xi_c$ molecule, respectively, and the coupled channel effects are not included.
It can be found that for the case of $\Lambda_{c\bar{c}}(4213)$ as a $D_s^{-} \Lambda_c^+$ molecule, the primary decay channel is $\bar{K}N$, which is about half of the total decay width, and this conclusion is the same as in Ref.~\cite{Wu:2010jy}.
However, the numerical decay width value of $\bar{K}N$ is about 4 MeV, which is much smaller than that in Ref.~\cite{Wu:2010jy} with about a factor 4.
Furthermore, for other light pseudoscalar meson light baryon decay channels, the decay widths in our model are all smaller with a factor 3 to 4.
The coupled channel effects do have effects on the decay ratios, but it is not a severe problem here since the overall difference could be removed by resetting the values of two cut-offs or the couplings.
In addition, both in these two models, $\eta_c \Lambda$ is the secondary dominating decay channel, which occupies around 20\% of the total decay width.
%
In summary, we can see from the comparison above that the coupled channel effects will not influence the decay estimation results heavily, thus the results calculated through the triangle diagrams are reasonable.
Through these comparison, it implies that $\eta_c \Lambda$ and $\bar{K} N$ channels could be the appropriate ones to search for the $\Lambda_{c\bar{c}}(4213)$ state.

\begin{table}
\caption{\label{table:DsLc12}{The dependence of the partial decay widths of $\Lambda_{c \bar{c}}(4213)$ as a $J^P=\frac12^-$ $D_s^{-} \Lambda_c^+$ molecule to $\eta_c \Lambda$, $\bar{K} N$ and $\bar K^* N$ channels on the cutoffs: (upper) $\Lambda_2$ changes with $\Lambda_1$ fixed at 1.0~GeV; (nether) $\Lambda_1$ changes with $\Lambda_2$ fixed at 2.0~GeV.} }
\renewcommand\arraystretch{1.2}
\begin{tabular}{c|*{3}{p{1.7cm}<{\centering}}}
\hline
& \multicolumn{3}{c}{Widths ($\mathrm{MeV}$) with $\Lambda_1=1.0~\mathrm{GeV}$} \\
\hline
$\Lambda_2$ ($\mathrm{GeV}$)    & 1.8   & 2.0   &	2.2	  \\
\hline
$\eta_c \Lambda$                & 1.736 & 2.624 &	3.584 \\
\hline
$\bar{K} N$                     & 2.036 &	4.417 &	8.157	\\
\hline
$\bar K^* N$                    & 0.756 &	1.301 & 2.095 \\
\hline
\end{tabular}
\begin{tabular}{c|*{3}{p{1.7cm}<{\centering}}}
\hline
& \multicolumn{3}{c}{Widths ($\mathrm{MeV}$) with $\Lambda_2=2.0~\mathrm{GeV}$} \\
\hline
$\Lambda_1$ ($\mathrm{GeV}$)    & 0.8   & 1.0   &	1.2	  \\
\hline
$\eta_c \Lambda$                & 1.446 & 2.624 &	4.170 \\
\hline
$\bar{K} N$                     & 2.268 &	4.417 &	7.604	\\
\hline
$\bar K^* N$                    & 0.682 &	1.301 & 2.199 \\
\hline
\end{tabular}
\end{table}

\begin{table}
\caption{\label{table:DXc12} {The dependence of the partial decay widths of $\Lambda_{c \bar{c}}(4213)$ as a $J^P=\frac12^-$ $\bar{D} \Xi_c$ molecule to $\eta_c \Lambda$, $\rho \Sigma$, $\pi \Sigma$ and $K \Xi$ channels on the cutoffs: (upper) $\Lambda_2$ changes with $\Lambda_1$ fixed at 1.0~GeV; (nether) $\Lambda_1$ changes with $\Lambda_2$ fixed at 2.0~GeV.} }
\renewcommand\arraystretch{1.2}
\begin{tabular}{c|*{3}{p{1.7cm}<{\centering}}}
\hline
& \multicolumn{3}{c}{Widths ($\mathrm{MeV}$) with $\Lambda_1=1.0~\mathrm{GeV}$} \\
\hline
$\Lambda_2$ ($\mathrm{GeV}$)    & 1.8   & 2.0   &	2.2	  \\
\hline
$\eta_c \Lambda$                & 1.428 &	1.961 &	2.478	\\
\hline
$\rho \Sigma$                   & 0.791 &	1.245 &	1.879	\\
\hline
$\pi \Sigma$                    & 1.238 &	2.902 &	5.553	\\
\hline
$K \Xi$                         & 0.650 &	1.528 & 2.941 \\
\hline
\end{tabular}
\begin{tabular}{c|*{3}{p{1.7cm}<{\centering}}}
\hline
& \multicolumn{3}{c}{Widths ($\mathrm{MeV}$) with $\Lambda_2=2.0~\mathrm{GeV}$} \\
\hline
$\Lambda_1$ ($\mathrm{GeV}$)    & 0.8   & 1.0   &	1.2	  \\
\hline
$\eta_c \Lambda$                & 0.965 &	1.961 &	3.390	\\
\hline
$\rho \Sigma$                   & 0.568 &	1.245 &	2.338	\\
\hline
$\pi \Sigma$                    & 1.308 &	2.902 &	5.540	\\
\hline
$K \Xi$                         & 0.691 &	1.528 & 2.905 \\
\hline
\end{tabular}
\end{table}

The S-wave $\bar{D} \Xi_c^\prime$ state named as $\Lambda_{c \bar{c}}(4403)$ is considered with $J^P=\frac12^-$.
Among all possible decay channels, $\eta_c \Lambda$ is the most important, since it provides more than 70\% to the total decay width, which is about 12.6~MeV.
The secondary and tertiary dominating decay channels are $\pi \Sigma$ and $\eta^\prime \Lambda$.
Similarly, the dependence of the decay widths of these three channels on the cutoffs $\Lambda_1$ and $\Lambda_2$ are calculated.
In Table~\ref{table:DXcpm12}, the numerical results are given.
This result is also smaller than that in Ref.~\cite{Wu:2010jy}, although the partial widths of the largest decay channel $\eta_c \Lambda$ is very similar around 10-15~MeV.
The main difference is due to the very small partial widths of $\pi \Sigma$ and $\eta^\prime \Lambda$ from our calculation.
Actually, in the reactions $\bar{D}\Xi_c^\prime \to \pi \Sigma$ or $\eta^\prime \Lambda$, one exchanges a deep off-shell $D^*$ particle, therefore, the amplitude strongly suffers from the form factor formalism and corresponding cutoffs.
Obviously, the cutoff regularization in Ref.~\cite{Wu:2010jy} is very different from the one in the present work, thus the partial width predictions have strong model dependence for such cases.

\begin{table}
\caption{\label{table:DXcpm12}{The dependence of the partial decay widths of $\Lambda_{c \bar{c}}(4403)$ as a $J^P=\frac12^-$ $\bar{D} \Xi_c^\prime$ molecule to $\eta_c \Lambda$, $\pi \Sigma$ and $\eta^\prime \Lambda$ channels on the cutoffs: (upper) $\Lambda_2$ changes with $\Lambda_1$ fixed at 1.0~GeV; (nether) $\Lambda_1$ changes with $\Lambda_2$ fixed at 2.0~GeV.} }
\renewcommand\arraystretch{1.2}
\begin{tabular}{c|*{3}{p{1.7cm}<{\centering}}}
\hline
& \multicolumn{3}{c}{Widths ($\mathrm{MeV}$) with $\Lambda_1=1.0~\mathrm{GeV}$} \\
\hline
$\Lambda_2$ ($\mathrm{GeV}$)    & 1.8   & 2.0   &	2.2	  \\
\hline
$\eta_c \Lambda$                & 6.482 &	9.126 &	11.751	\\
\hline
$\pi \Sigma$                    & 0.441 &	1.008 &	1.921	\\
\hline
$\eta^\prime \Lambda$           & 0.409 &	0.845 & 1.515 \\
\hline
\end{tabular}
\begin{tabular}{c|*{3}{p{1.7cm}<{\centering}}}
\hline
& \multicolumn{3}{c}{Widths ($\mathrm{MeV}$) with $\Lambda_2=2.0~\mathrm{GeV}$} \\
\hline
$\Lambda_1$ ($\mathrm{GeV}$)    & 0.8   & 1.0   &	1.2	  \\
\hline
$\eta_c \Lambda$                & 4.909 &	9.126 &	14.773	\\
\hline
$\pi \Sigma$                    & 0.518 &	1.008 &	1.737	\\
\hline
$\eta^\prime \Lambda$           & 0.427 &	0.845 & 1.474 \\
\hline
\end{tabular}
\end{table}

One sees that the total decay width of the $\Lambda_{c \bar{c}}(4370)$ state described as a $J^P=\frac12^-$ $D_s^{* -} \Lambda_c^+$ molecule is 32~MeV with $\Lambda_1=1.0\ \mathrm{GeV}$ and $\Lambda_2=2.0\ \mathrm{GeV}$.
This value is much larger, by one order of magnitude, than that 2.8~MeV with $J^P=\frac32^-$.
We found that the $D^*$ exchange in $\bar K^* N$ channel contributes the most to total decay width in both spin-parity-$1/2^-$ and $3/2^-$ cases.
The first three dominant two body decay channels of $\Lambda_{c \bar{c}}(4370)$ are $J/\psi \Lambda$, $\phi \Lambda$ and $\bar K^* N$ in this $D_s^{* -} \Lambda_c^+$ hadronic molecular assumption.
Since the branching ratio of these three decay channels have already reached 89\% and 95\% for $J^P=\frac12^-$ and $J^P=\frac32^-$ cases, respectively, we will discuss the dependence of the decay width on the cutoffs only in these three channels.
These dependence results are given in Tables~\ref{table:DssLc12} and \ref{table:DssLc32}.

\begin{table}
\caption{\label{table:DssLc12}The dependence of the partial decay widths of $\Lambda_{c \bar{c}}(4370)$ as a $J^P=\frac12^-$ $D_s^{* -} \Lambda_c^+$ molecule to $J/\psi \Lambda$, $\phi \Lambda$ and $\bar K^* N$ channels on the cutoffs: (upper) $\Lambda_2$ changes with $\Lambda_1$ fixed at 1.0~GeV; (nether) $\Lambda_1$ changes with $\Lambda_2$ fixed at 2.0~GeV.}
\renewcommand\arraystretch{1.2}
\begin{tabular}{c|*{3}{p{1.7cm}<{\centering}}}
\hline
& \multicolumn{3}{c}{Widths ($\mathrm{MeV}$) with $\Lambda_1=1.0~\mathrm{GeV}$} \\
\hline
$\Lambda_2$ ($\mathrm{GeV}$)    & 1.8   & 2.0   &	2.2	  \\
\hline
$J/\psi \Lambda$                & 1.919 & 3.006 &	4.217 \\
\hline
$\phi \Lambda$                  & 2.499 &	4.403 &	7.270	\\
\hline
$\bar K^* N$                    & 11.997 &	20.945 & 34.237 \\
\hline
\end{tabular}
\begin{tabular}{c|*{3}{p{1.7cm}<{\centering}}}
\hline
& \multicolumn{3}{c}{Widths ($\mathrm{MeV}$) with $\Lambda_2=2.0~\mathrm{GeV}$} \\
\hline
$\Lambda_1$ ($\mathrm{GeV}$)    & 0.8   & 1.0   &	1.2	  \\
\hline
$J/\psi \Lambda$                & 1.679 & 3.006 &	4.692 \\
\hline
$\phi \Lambda$                  & 2.159 &	4.403 &	7.968	\\
\hline
$\bar K^* N$                    & 10.226 &	20.945 & 38.058 \\
\hline
\end{tabular}
\end{table}

\begin{table}
\caption{\label{table:DssLc32}The dependence of the partial decay widths of $\Lambda_{c \bar{c}}(4370)$ as a $J^P=\frac32^-$ $D_s^{* -} \Lambda_c^+$ molecule to $J/\psi \Lambda$, $\phi \Lambda$ and $\bar K^* N$ channels on the cutoffs: (upper) $\Lambda_2$ changes with $\Lambda_1$ fixed at 1.0~GeV; (nether) $\Lambda_1$ changes with $\Lambda_2$ fixed at 2.0~GeV.}
\renewcommand\arraystretch{1.2}
\begin{tabular}{c|*{3}{p{1.7cm}<{\centering}}}
\hline
& \multicolumn{3}{c}{Widths ($\mathrm{MeV}$) with $\Lambda_1=1.0~\mathrm{GeV}$} \\
\hline
$\Lambda_2$ ($\mathrm{GeV}$)    & 1.8   & 2.0   &	2.2	  \\
\hline
$J/\psi \Lambda$                & 0.510 & 0.766 &	1.044 \\
\hline
$\phi \Lambda$                  & 0.248 &	0.380 &	0.572	\\
\hline
$\bar K^* N$                    & 1.023 &	1.527 & 2.247 \\
\hline
\end{tabular}
\begin{tabular}{c|*{3}{p{1.7cm}<{\centering}}}
\hline
& \multicolumn{3}{c}{Widths ($\mathrm{MeV}$) with $\Lambda_2=2.0~\mathrm{GeV}$} \\
\hline
$\Lambda_1$ ($\mathrm{GeV}$)    & 0.8   & 1.0   &	1.2	  \\
\hline
$J/\psi \Lambda$                & 0.429 & 0.766 &	1.214 \\
\hline
$\phi \Lambda$                  & 0.163 &	0.380 &	0.796	\\
\hline
$\bar K^* N$                    & 0.639 &	1.527 & 3.281 \\
\hline
\end{tabular}
\end{table}

The $\Lambda_{c \bar{c}}(4370)$ represents a pair of degenerate $\bar{D}^{*} \Xi_c$ bound states with spin-parity-$1/2^-$ and $3/2^-$, respectively.
The total decay width with $J^P=\frac12^-$ is about 42~MeV, which is much larger than the width 3.5~MeV with $J^P=\frac32^-$.
$J/\psi \Lambda$, $\rho \Sigma$, $\omega \Lambda$ and $K^* \Xi$ are the four primary channels of $\Lambda_{c \bar{c}}(4370)$ with $J^P=\frac32^-$, while two more channels $\bar{D} \Xi_c$ and $D^-_s \Lambda^+_c$ are needed when discussing the main decay channels with $J^P=\frac12^-$.
The decay widths of these channels occupy more than 96\% of the total decay width in both two cases.
Among all the final states considered, $\rho \Sigma$ with $D^*$ exchange dominates, followed by $K^* \Xi$ with $D_s^*$ exchange.
In Tables~\ref{table:DstXc12} and \ref{table:DstXc32}, we display the partial decay widths dependence on $\Lambda_1$ and $\Lambda_2$ of these decay channels.

\begin{table}
\caption{\label{table:DstXc12}The dependence of the partial decay widths of $\Lambda_{c \bar{c}}(4370)$ as a $J^P=\frac12^-$ $\bar{D}^{*} \Xi_c$ molecule to $J/\psi \Lambda$, $\bar{D} \Xi_c$, $D^-_s \Lambda^+_c$, $\rho \Sigma$, $\omega \Lambda$ and $K^* \Xi$ channels on the cutoffs: (upper) $\Lambda_2$ changes with $\Lambda_1$ fixed at 1.0~GeV; (nether) $\Lambda_1$ changes with $\Lambda_2$ fixed at 2.0~GeV.}
\renewcommand\arraystretch{1.2}
\begin{tabular}{c|*{3}{p{1.7cm}<{\centering}}}
\hline
& \multicolumn{3}{c}{Widths ($\mathrm{MeV}$) with $\Lambda_1=1.0~\mathrm{GeV}$} \\
\hline
$\Lambda_2$ ($\mathrm{GeV}$)    & 1.8   & 2.0   &	2.2	  \\
\hline
$J/\psi \Lambda$                & 1.660 &	2.380 &	3.106	\\
\hline
$\bar{D} \Xi_c$                 & 2.636 &	3.171 &	3.594	\\
\hline
$D^-_s \Lambda^+_c$             & 3.024 &	3.862 &	4.565	\\
\hline
$\rho \Sigma$                   & 13.179 &	20.696 &	31.423	\\
\hline
$\omega \Lambda$                & 1.467 &	2.304 &	3.503	\\
\hline
$K^* \Xi$                       & 5.352 &	8.463 & 12.992 \\
\hline
\end{tabular}
\begin{tabular}{c|*{3}{p{1.7cm}<{\centering}}}
\hline
& \multicolumn{3}{c}{Widths ($\mathrm{MeV}$) with $\Lambda_2=2.0~\mathrm{GeV}$} \\
\hline
$\Lambda_1$ ($\mathrm{GeV}$)    & 0.8   & 1.0   &	1.2	  \\
\hline
$J/\psi \Lambda$                & 1.164 &	2.380 &	4.096	\\
\hline
$\bar{D} \Xi_c$                 & 1.387 &	3.171 &	5.466	\\
\hline
$D^-_s \Lambda^+_c$             & 1.815 &	3.862 &	6.376	\\
\hline
$\rho \Sigma$                   & 8.418 &	20.696 &	43.220	\\
\hline
$\omega \Lambda$                & 0.936 &	2.304 &	4.817	\\
\hline
$K^* \Xi$                       & 3.454 &	8.463 & 17.626 \\
\hline
\end{tabular}
\end{table}

\begin{table}
\caption{\label{table:DstXc32}The dependence of the partial decay widths of $\Lambda_{c \bar{c}}(4370)$ as a $J^P=\frac32^-$ $\bar{D}^{*} \Xi_c$ molecule to $J/\psi \Lambda$, $\rho \Sigma$, $\omega \Lambda$ and $K^* \Xi$ channels on the cutoffs: (upper) $\Lambda_2$ changes with $\Lambda_1$ fixed at 1.0~GeV; (nether) $\Lambda_1$ changes with $\Lambda_2$ fixed at 2.0~GeV.}
\renewcommand\arraystretch{1.2}
\begin{tabular}{c|*{3}{p{1.7cm}<{\centering}}}
\hline
& \multicolumn{3}{c}{Widths ($\mathrm{MeV}$) with $\Lambda_1=1.0~\mathrm{GeV}$} \\
\hline
$\Lambda_2$ ($\mathrm{GeV}$)    & 1.8   & 2.0   &	2.2	  \\
\hline
$J/\psi \Lambda$                & 0.436 &	0.593 &	0.744	\\
\hline
$\rho \Sigma$                   & 1.362 &	1.781 &	2.355	\\
\hline
$\omega \Lambda$                & 0.148 &	0.193 &	0.255	\\
\hline
$K^* \Xi$                       & 0.611 &	0.824 & 1.123 \\
\hline
\end{tabular}
\begin{tabular}{c|*{3}{p{1.7cm}<{\centering}}}
\hline
& \multicolumn{3}{c}{Widths ($\mathrm{MeV}$) with $\Lambda_2=2.0~\mathrm{GeV}$} \\
\hline
$\Lambda_1$ ($\mathrm{GeV}$)    & 0.8   & 1.0   &	1.2	  \\
\hline
$J/\psi \Lambda$                & 0.289 &	0.593 &	1.040	\\
\hline
$\rho \Sigma$                   & 0.607 &	1.781 &	4.387	\\
\hline
$\omega \Lambda$                & 0.066 &	0.193 &	0.478	\\
\hline
$K^* \Xi$                       & 0.289 &	0.824 & 1.982 \\
\hline
\end{tabular}
\end{table}

In Ref.~\cite{Wu:2010jy}, Wu {\it et~al.} estimate the decay widths of the predicted $\Lambda_{c \bar{c}}(4370)$ states from $VB \to VB$ interaction by exchanging a vector meson.
The total decay width of $\Lambda_{c \bar{c}}(4370)$ is 28.0~MeV with 13.9~MeV from $\bar K^* N$, 3.1~MeV from $\rho \Sigma$, 0.3~MeV from $\omega \Lambda$, 4.0~MeV from $\phi \Lambda$, 1.8~MeV from $K^* \Xi$ and 5.4~MeV from $J/\psi \Lambda$.
Since $\Lambda_{c \bar{c}}(4370)$ couples to both $D_s^{* -} \Lambda_c^+$ and $\bar{D}^{*} \Xi_c$ in Ref.~\cite{Wu:2010jy}, the fourth(fifth) and sixth(seventh) column in Table~\ref{table:width} should be combined when comparing these two partial decay widths.
We can find that the decay patterns with $J^P=\frac12^-$ is in good accordance with the results in Ref.~\cite{Wu:2010jy}.
The difference comes from that in our calculation both pseudoscalar and vector meson exchange are involved, while in Ref.~\cite{Wu:2010jy} only vector meson exchange is considered.
Therefore, if there is only one $\Lambda_{c \bar{c}}$ state in this energy range, it can be distinguished by the value of the total width, i.e., $\Lambda_{c \bar{c}}(4370)(\frac{1}{2}^-)$ state prefer a broad state, while $J^P=\frac32^-$ state will be very narrow.
%

The $\Lambda_{c\bar{c}}(4550)$ represents a pair of degenerate $\bar{D}^* \Xi_c^\prime$ bound states predicted in Ref.~\cite{Wu:2010jy} for spin-parity of $1/2^-$ and $3/2^-$, respectively.
The three dominating decay channels are $\bar D \Xi_c$, $\bar D^* \Xi_c$ and $\bar D \Xi_c^\prime$ for $J^P={\frac12}^-$ case and $\bar D \Xi_c$, $\bar D^* \Xi_c$ and $\bar D \Xi_c^*$ for $J^P={\frac32}^-$ case.
The dependence of the partial decay widths of these main channels on cutoffs in both two cases are listed in Tables~\ref{table:DstXcpm12} and \ref{table:DstXcpm32}.
The total decay width of $\Lambda_{c \bar{c}}(4550)$ is 36.6~MeV in Ref.~\cite{Wu:2010jy}, and only $\rho \Sigma$, $\omega \Lambda$, $K^* \Xi$ and $J/\psi \Lambda$ four channels are considered in the coupled channel calculation.
From our results, the values of partial widths of these four dominant channels in Ref.~\cite{Wu:2010jy} are well consistent in $J^P={\frac12}^-$ case.
However, in our calculation the dominant decay channels are charmed baryon and anticharmed meson channels, $\bar D \Xi_c$, $\bar D^* \Xi_c$ and $\bar D \Xi_c^\prime$, and their decay widths are larger than others by one magnitude order.
The main reason is that these reactions exchange light pseudoscalar and vector meson, while in Ref.~\cite{Wu:2010jy} they missed various interaction vertices, such as $VVP$, $BBP$, $DBP$ and $DBV$ vertices.
This conclusion is in accordance with the results of the $P_c$ states in Ref.~\cite{Shen:2016tzq,Lin:2017mtz}, where the dominating decay channel of $P_c$ is $\bar{D}^{(*)}\Lambda_c$.
According to our calculation, the total decay width of $\Lambda_{c \bar{c}}(4550)$ with $J^P={\frac32}^-$ is about 40~MeV, while in the $J^P={\frac12}^-$ case, the width of $\Lambda_{c \bar{c}}(4550)$ is quite broad, since both $\bar D \Xi_c$ and $\bar D \Xi_c^\prime$ final states have a width of about 100~MeV.

\begin{table}
\caption{\label{table:DstXcpm12}The dependence of the partial decay widths of $\Lambda_{c \bar{c}}(4550)$ as a $J^P=\frac12^-$ $\bar{D}^* \Xi_c^\prime$ molecule to $\bar D \Xi_c$, $\bar D^* \Xi_c$ and $\bar D \Xi_c^\prime$ channels on the cutoffs: (upper) $\Lambda_2$ changes with $\Lambda_1$ fixed at 1.0~GeV; (nether) $\Lambda_1$ changes with $\Lambda_2$ fixed at 2.0~GeV.}
\renewcommand\arraystretch{1.2}
\begin{tabular}{c|*{3}{p{1.7cm}<{\centering}}}
\hline
& \multicolumn{3}{c}{Widths ($\mathrm{MeV}$) with $\Lambda_1=1.0~\mathrm{GeV}$} \\
\hline
$\Lambda_2$ ($\mathrm{GeV}$)    & 1.8   & 2.0   &	2.2	  \\
\hline
$\bar D \Xi_c$                  & 89.808 & 99.249 &	106.417 \\
\hline
$\bar D^* \Xi_c$                & 47.464 &	50.646 &	52.942	\\
\hline
$\bar D \Xi_c^\prime$           & 97.676 &	106.064 & 112.303 \\
\hline
\end{tabular}
\begin{tabular}{c|*{3}{p{1.7cm}<{\centering}}}
\hline
& \multicolumn{3}{c}{Widths ($\mathrm{MeV}$) with $\Lambda_2=2.0~\mathrm{GeV}$} \\
\hline
$\Lambda_1$ ($\mathrm{GeV}$)    & 0.8   & 1.0   &	1.2	  \\
\hline
$\bar D \Xi_c$                  & 56.248 & 99.249 &	149.577 \\
\hline
$\bar D^* \Xi_c$                & 29.299 &	50.646 &	73.882	\\
\hline
$\bar D \Xi_c^\prime$           & 58.174 &	106.064 & 163.452 \\
\hline
\end{tabular}
\end{table}

\begin{table}
\caption{\label{table:DstXcpm32}The dependence of the partial decay widths of $\Lambda_{c \bar{c}}(4550)$ as a $J^P=\frac32^-$ $\bar{D}^* \Xi_c^\prime$ molecule to $\bar D \Xi_c$, $\bar D^* \Xi_c$ and $\bar D \Xi_c^*$ channels on the cutoffs: (upper) $\Lambda_2$ changes with $\Lambda_1$ fixed at 1.0~GeV; (nether) $\Lambda_1$ changes with $\Lambda_2$ fixed at 2.0~GeV.}
\renewcommand\arraystretch{1.2}
\begin{tabular}{c|*{3}{p{1.7cm}<{\centering}}}
\hline
& \multicolumn{3}{c}{Widths ($\mathrm{MeV}$) with $\Lambda_1=1.0~\mathrm{GeV}$} \\
\hline
$\Lambda_2$ ($\mathrm{GeV}$)    & 1.8   & 2.0   &	2.2	  \\
\hline
$\bar D \Xi_c$                  & 4.593 & 5.050 &	5.398 \\
\hline
$\bar D^* \Xi_c$                & 4.716 &	5.025 &	5.248	\\
\hline
$\bar D \Xi_c^*$                & 21.602 &	24.555 & 26.816 \\
\hline
\end{tabular}
\begin{tabular}{c|*{3}{p{1.7cm}<{\centering}}}
\hline
& \multicolumn{3}{c}{Widths ($\mathrm{MeV}$) with $\Lambda_2=2.0~\mathrm{GeV}$} \\
\hline
$\Lambda_1$ ($\mathrm{GeV}$)    & 0.8   & 1.0   &	1.2	  \\
\hline
$\bar D \Xi_c$                  & 4.156 & 5.050 &	5.538 \\
\hline
$\bar D^* \Xi_c$                & 3.123 &	5.025 &	7.045	\\
\hline
$\bar D \Xi_c^*$                & 12.279 &	24.555 & 39.465 \\
\hline
\end{tabular}
\end{table}

In the present work, besides the four $\Lambda_{c \bar{c}}$ states predicted in Ref.~\cite{Wu:2010jy}, there is an additional S-wave $\bar{D} \Xi_c^*$ state included.
We choose it to be $\Lambda_{c \bar{c}}(4490)$ with spin-parity-$3/2^-$, whose binding energy equals to 23~MeV.
Since the two $P_c$ states are regarded as $\bar{D} \Sigma_c^*$ and $\bar{D}^* \Sigma_c$ molecules~\cite{Lu:2016nnt, Shen:2016tzq,Lin:2017mtz, Chen:2015moa, He:2015cea,Roca:2015dva, Huang:2015uda} and the $K \Sigma^*$ and $K^* \Sigma$ are applied to explain the nature of $N(1875)$ and $N(2080)$~\cite{Lin:2018kcc}, it is quite nature that there exist the $\bar D \Xi_c^*$ bound state together with the $\bar D^* \Xi_c$ bound state.
The partial decay widths of $\Lambda_{c \bar{c}}(4490)$ are already given in Table~\ref{table:width}.
It can be seen that $\rho \Sigma$, $\omega \Lambda$ and $\bar D^* \Xi_c$ are the first three dominating channels and $\bar D^* \Xi_c$ channel with $\pi$ exchange contributing the most among all the final states.
The dependence of the partial decay widths of these three final states on cutoffs are given in Table~\ref{table:DXcst32}.

\begin{table}
\caption{\label{table:DXcst32}The dependence of the partial decay widths of $\Lambda_{c \bar{c}}(4490)$ as a $J^P=\frac32^-$ $\bar{D} \Xi_c^*$ molecule to $\rho \Sigma$, $\omega \Lambda$ and $\bar D^* \Xi_c$ channels on the cutoffs: (upper) $\Lambda_2$ changes with $\Lambda_1$ fixed at 1.0~GeV; (nether) $\Lambda_1$ changes with $\Lambda_2$ fixed at 2.0~GeV.}
\renewcommand\arraystretch{1.2}
\begin{tabular}{c|*{3}{p{1.7cm}<{\centering}}}
\hline
& \multicolumn{3}{c}{Widths ($\mathrm{MeV}$) with $\Lambda_1=1.0~\mathrm{GeV}$} \\
\hline
$\Lambda_2$ ($\mathrm{GeV}$)    & 1.8   & 2.0   &	2.2	  \\
\hline
$\rho \Sigma$                   & 1.691 & 2.838 &	4.548 \\
\hline
$\omega \Lambda$                & 1.691 &	2.824 &	4.516	\\
\hline
$\bar D^* \Xi_c$                & 14.956 &	15.709 & 16.243 \\
\hline
\end{tabular}
\begin{tabular}{c|*{3}{p{1.7cm}<{\centering}}}
\hline
& \multicolumn{3}{c}{Widths ($\mathrm{MeV}$) with $\Lambda_2=2.0~\mathrm{GeV}$} \\
\hline
$\Lambda_1$ ($\mathrm{GeV}$)    & 0.8   & 1.0   &	1.2	  \\
\hline
$\rho \Sigma$                   & 1.439 & 2.838 &	4.966 \\
\hline
$\omega \Lambda$                & 1.425 &	2.824 &	4.970	\\
\hline
$\bar D^* \Xi_c$                & 8.904 &	15.709 & 23.890 \\
\hline
\end{tabular}
\end{table}

As the $\Lambda_{c \bar{c}}(4490)$ state is not predicted in previous works, we should also discuss the dependence of its decay widths on the binding energy.
Since the thresholds of $\bar{D}^* \Xi_c$ is 4478~MeV, and the state should not be tightly bound, we only range the mass of the state from 4480 to 4500~MeV.
The numerical results are shown in Table~\ref{table:DXcstBE}.
It turns out that the partial decay width of the $\bar D^* \Xi_c$ channel relies the most on the binding energy, while the widths of other decay channels vary very little when the binding energy changes.
The reason could be that the state locates very close to the threshold of $\bar D^* \Xi_c$, the phase space influences the most for this channel especially when the $\bar{D} \Xi_c^*$ state is bounded tightly.
It can be seen that the branching ratio of $\bar D^* \Xi_c$ is always the largest, which is similar as what happens in the $P_c$'s case~\cite{Shen:2016tzq,Lin:2017mtz}.
And the binding energy being 23~MeV could be a proper choice when discussing this $\bar{D} \Xi_c^*$ bound state.
It is a firm conclusion that the total decay width of the possible $\bar{D} \Xi_c^*$ state should be around 25~MeV with $\bar D^* \Xi_c$ being the primary final state.

\begin{table}
\caption{\label{table:DXcstBE}The dependence of the decay widths of $\Lambda_{c \bar{c}}$ as a $J^P=\frac32^-$ $\bar{D} \Xi_c^*$ molecule to possible decay channels on the binding energy with $\Lambda_1=1.0\ \mathrm{GeV}$ and $\Lambda_2=2.0\ \mathrm{GeV}$.}
\renewcommand\arraystretch{1.43}
\begin{tabular}{l|*{3}{p{2.1cm}<{\centering}}}
\hline
\multirow{4}*{Mode} & \multicolumn{3}{c}{Widths ($\mathrm{MeV}$)} \\
\cline{2-4}
& \multicolumn{3}{c}{$\bar{D} \Xi_c^*$ ($J^P={\frac32}^-$)} \\
\cline{2-4}
& \multicolumn{1}{c}{$\Lambda_{c \bar{c}}(4480)$} & \multicolumn{1}{c}{$\Lambda_{c \bar{c}}(4490)$} & \multicolumn{1}{c}{$\Lambda_{c \bar{c}}(4500)$} \\
\hline
$J/\psi \Lambda$			&  0.453 &	0.464 &	0.453	\\
$\eta_c \Lambda$			&	 0.107 &	0.115 & 0.121	\\
$\rho \Sigma$				&  2.986  &	2.837 &	2.558	\\
$\omega \Lambda$			&  2.976  &	2.824 &	2.542	\\
$\pi \Sigma$				& 0.172 &	0.165 & 0.152	\\
$\eta \Lambda$				& 0.058  &	0.056 & 0.052	\\
$\eta^\prime \Lambda$		&  0.112 &	0.109 & 0.101	\\
$K \Xi$					& 0.078 &	0.075 &	0.069	\\
$K^* \Xi$					& 1.190 &	1.130 &	1.017	\\
$D_s^{* -} \Lambda_c^+$		&  1.219 &	1.304 &	1.334	\\
$\bar D^* \Xi_c$			&  6.302  & 15.709 &	20.365	\\
$\bar D \Xi_c^\prime$		& 0.012  &	0.024 &	0.042	\\
\hline
Total						& 15.665  &	24.812  &	28.806	\\
\hline
\end{tabular}
\end{table}

According to our calculation, we suggest to search for $\Lambda_{c \bar{c}}(4213)$ and $\Lambda_{c \bar{c}}(4403)$ states in the $\eta_c \Lambda$ system and $\Lambda_{c \bar{c}}(4490)$ and $\Lambda_{c \bar{c}}(4550)$ states in the charmed baryon and anticharmed meson system, eg. $\bar D^* \Xi_c$.
And it should be easier to search for $\Lambda_{c \bar{c}}(4370)$ state in the $\rho \Sigma$, $\bar{K}^* N$ or $K^* \Xi$ production than others.
It is quite meaningful to search for these $\Lambda_{c \bar{c}}$ states experimentally and the experimental results will strongly help to disentangle the nature of these $\Lambda_{c \bar{c}}$ structures.

\section{Summary} \label{sec:summary}

Inspired by the success of the investigation on the decay behaviors of $P_c$ states, we extend the approach to study the two body decays of $\Lambda_{c \bar{c}}$ states through triangle diagram with mesons exchange in different hadronic molecular assumptions.
According to Ref.~\cite{Wu:2010jy}, the predicted six $\Lambda_{c \bar{c}}$ states can be divided into two groups, two pseudoscalar meson baryon molecules and four vector meson baryon molecules.
In various anticharmed meson and charmed baryon molecular assumptions with $J^P=\frac12^-$ or $\frac32^-$, the possible partial decay widths are calculated.
For the two pseudoscalar meson baryon molecule $\Lambda_{c\bar{c}}$ states, their $J^P$ only can be $\frac12^-$ for S-wave interaction.
$\Lambda_{c \bar{c}}(4213)$ could be either a $D_s^{-} \Lambda_c^+$ or $\bar{D} \Xi_c$ bound state, and the total decay width in these two assumptions are similar, which is around 10~MeV.
But the main decay channels in these two cases are different.
For $D_s^{-} \Lambda_c^+$ molecule, the main decay channels are $\bar{K}N$ and $\eta_c \Lambda$, while it is $\pi \Sigma$ and $\eta_c \Lambda$ channels for $\bar{D}\Xi_c$ molecule.
$\Lambda_{c \bar{c}}(4403)$ is an S-wave $\bar{D} \Xi_c^\prime$ molecular state, whose dominant decay channel is $\eta_c \Lambda$ and its decay width is consistent with that in Ref.~\cite{Wu:2010jy}.
However, for other two important decay channels without $c\bar{c}$ components, $\pi \Sigma$ and $\eta \Lambda$, the decay widths are much smaller than that in Ref.~\cite{Wu:2010jy}.
It suggests that such widths suffer from model dependence because the exchanged particle could be far off-shell in some cases.
In summary, for these two $\Lambda_{c \bar{c}}$ states, the total decay widths are dozens of MeV, and $\eta_c \Lambda$ is the best channel to search for them.
The other four $\Lambda_{c \bar{c}}$ states are formed by a vector meson and a baryon, thus their $J^P$ can be $\frac12^-$ or $\frac32^-$.
According to our calculation, the total width with $J^P=\frac32^-$ is much smaller than that with $J^P=\frac12^-$.
$\Lambda_{c \bar{c}}(4370)$ could either be a $D_s^{* -} \Lambda_c^+$ or $\bar{D}^{*} \Xi_c$ bound state.
We find that when $\Lambda_{c \bar{c}}(4370)$ is a $D_s^{* -} \Lambda_c^+$ molecule, $J/\psi \Lambda$, $\phi \Lambda$ and $\bar K^* N$ are the three dominating final states, while $\rho \Sigma$ and $K^* \Xi$ occupy more for $\Lambda_{c \bar{c}}(4370)$ being a $\bar{D}^{*} \Xi_c$ bound state.
The decay patterns we get with $J^P=\frac12^-$ are in good accordance with the predicted results in Ref.~\cite{Wu:2010jy}.
The predicted two nearly degenerate $\Lambda_{c \bar{c}}$ states around 4550~MeV with $J^P$ of either $1/2^-$ or $3/2^-$ are $\bar D^* \Xi_c^\prime$ molecular states.
Since the light pseudoscalar mesons exchange are included in this work, it can decay to anticharmed meson and charmed baryon final states, such as $\bar{D}^{(*)} \Xi_c $, $\bar{D}\Xi_c^{*}$ and $\bar{D}\Xi_c^\prime$.
This leads to a very broad width with $J^P=\frac12^-$, which is about 300~MeV, while the width is about 40~MeV with $J^P=\frac32^-$.
This result is consistent with the conclusion in the $P_c$'s case, where the main decay channels of $P_c$ are $\bar{D}^{(*)}\Lambda_c$.
Thus, for such states, the $\bar{D}^{(*)} \Xi^*_c $ channels should be good choices to search for.

At last, an additional S-wave $\bar{D} \Xi_c^*$ bound state is supposed to exist in the calculation, which is $\Lambda_{c \bar{c}}(4490)$.
The results tell that its primary decay channels are $\rho \Sigma$, $\omega \Lambda$ and $\bar D^* \Xi_c$.
The dependence of partial decay widths on the binding energy is also discussed.
And it is found that the total width is rather stable when the binding energy varies, which is always between 10 to 30~MeV.

In summary, we have studied the decay behaviors of seven $\Lambda_{c \bar{c}}$ states.
Different molecular assumptions and spin-parities will lead to very different total widths and decay patterns.
It is quite promising to find these states experimentally in the future.

\section*{Acknowledgments}

We thank the fruitful discussion with Jujun Xie.
This project is supported by NSFC under Grant No.~11621131001 (CRC110 cofunded by DFG and NSFC) and Grant No.~11835015.
This project is also supported by the Thousand Talents Plan for Young Professionals.

\begin{appendix}

  \section{couplings of triangle diagram}\label{AppendixA}
  The coupling constants of the vertices in the triangle diagrams are related to each other by SU(4) flavor symmetry~\cite{Okubo:1975sc,Liu:2001ce,Dong:2009tg} or heavy quark symmetry, chiral symmetry and hidden local symmetry~\cite{Guo:2010ak,Chen:2018pzd,Wang:2018cre}. The employed relations for various coupling constants are given by the following expressions:
  \begin{eqnarray}
  g_{\Lambda_c N D} &=& -\frac{3\sqrt3}5 g_{BBP}, \nonumber \\
  g_{\Lambda_c \Lambda D_s} &=& \frac{3\sqrt2}5 g_{BBP}, \nonumber \\
  g_{\Lambda_c \Lambda_c \eta_c} &=& \frac{3\sqrt2}5 g_{BBP}, \nonumber \\
  g_{\Xi_c \Lambda D} &=& -\frac3{5\sqrt2} g_{BBP}, \nonumber \\
  g_{\Xi_c \Sigma D} &=& \frac{3\sqrt3}5 g_{BBP}, \nonumber \\
  g_{\Xi_c \Xi D_s} &=& -\frac{3\sqrt3}5 g_{BBP}, \nonumber \\
  g_{\Xi_c \Xi_c \eta_c} &=& \frac{3\sqrt2}5 g_{BBP}, \nonumber \\
  g_{\Xi_c^\prime \Lambda D} &=& \frac{\sqrt3}{5\sqrt2} g_{BBP}, \nonumber \\
  g_{\Xi_c^\prime \Sigma D} &=& \frac15 g_{BBP}, \nonumber \\
  g_{\Xi_c^\prime \Xi D_s} &=& \frac15 g_{BBP}, \nonumber \\
  g_{\Xi_c^\prime \Lambda_c K} &=& -\frac{\sqrt6}5 g_{BBP}, \nonumber \\
  g_{\Xi_c^\prime \Xi_c \pi} &=& -\frac{\sqrt6}5 g_{BBP}, \nonumber \\
  g_{\Xi_c^\prime \Xi_c \eta} &=& -\frac35 g_{BBP}, \nonumber \\
  g_{\Xi_c^\prime \Xi_c^\prime \pi} &=& \frac{2\sqrt2}5 g_{BBP}, \nonumber \\
  g_{\Xi_c^\prime \Xi_c^\prime \eta} &=& -\frac2{5\sqrt3} g_{BBP}, \nonumber \\
  g_{\Xi_c^\prime \Xi_c^\prime \eta^\prime} &=& \frac{4\sqrt2}{5\sqrt3} g_{BBP}, \nonumber \\
  g_{\Lambda_c N D^*} &=& -\sqrt3 g_{BBV}, \nonumber \\
  g_{\Lambda_c \Lambda D_s^*} &=& \sqrt2 g_{BBV}, \nonumber \\
  g_{\Lambda_c \Lambda_c \omega} &=& 2 g_{BBV}, \nonumber \\
  g_{\Lambda_c \Lambda_c J/\psi} &=& \sqrt2 g_{BBV}, \nonumber \\
  g_{\Lambda_c \Xi_c K^*} &=& \sqrt2 g_{BBV}, \nonumber \\
  g_{\Xi_c \Lambda D^*} &=& -\frac1{\sqrt2} g_{BBV}, \nonumber \\
  g_{\Xi_c \Sigma D^*} &=& \sqrt3 g_{BBV}, \nonumber \\
  g_{\Xi_c \Xi D_s^*} &=& -\sqrt3 g_{BBV}, \nonumber \\
  g_{\Xi_c \Lambda_c K^*} &=& -\sqrt2 g_{BBV}, \nonumber \\
  g_{\Xi_c \Xi_c \rho} &=& \sqrt2 g_{BBV}, \nonumber \\
  g_{\Xi_c \Xi_c \omega} &=& g_{BBV}, \nonumber \\
  g_{\Xi_c \Xi_c \phi} &=& -\sqrt2 g_{BBV}, \nonumber \\
  g_{\Xi_c \Xi_c J/\psi} &=& \sqrt2 g_{BBV}, \nonumber \\
  g_{\Xi_c^\prime \Lambda D^*} &=& -\sqrt{\frac32} g_{BBV}, \nonumber \\
  g_{\Xi_c^\prime \Sigma D^*} &=& - g_{BBV}, \nonumber \\
  g_{\Xi_c^\prime \Xi D_s^*} &=& - g_{BBV}, \nonumber \\
  g_{\Xi_c^\prime \Xi_c^\prime \rho} &=& \sqrt2 g_{BBV}, \nonumber \\
  g_{\Xi_c^\prime \Xi_c^\prime \omega} &=& g_{BBV}, \nonumber \\
  g_{\Xi_c^\prime \Xi_c^\prime \phi} &=& -\sqrt2 g_{BBV}, \nonumber \\
  g_{\Xi_c^\prime \Xi_c^\prime J/\psi} &=& \sqrt2 g_{BBV}, \nonumber \\
  g_{D_s^* K D} &=& \sqrt2 g_{PPV}, \nonumber \\
  g_{D_s^* \eta D_s} &=& -\frac2{\sqrt3} g_{PPV}, \nonumber \\
  g_{D_s^* \eta^\prime D_s} &=& \sqrt{\frac23} g_{PPV}, \nonumber \\
  g_{D_s^* \eta_c D_s} &=& -\sqrt2 g_{PPV}, \nonumber \\
  g_{D^* \pi D} &=& \sqrt2 g_{PPV}, \nonumber \\
  g_{D^* K D_s} &=& \sqrt2 g_{PPV}, \nonumber \\
  g_{D^* \eta D} &=& \frac1{\sqrt3} g_{PPV}, \nonumber \\
  g_{D^* \eta^\prime D} &=& \sqrt{\frac23} g_{PPV}, \nonumber \\
  g_{D^* \eta_c D} &=& -\sqrt2 g_{PPV}, \nonumber \\
  g_{D D \rho} &=& \sqrt2 g_{PPV}, \nonumber \\
  g_{D D \omega} &=& g_{PPV}, \nonumber \\
  g_{D D J/\psi} &=& -\sqrt2 g_{PPV}, \nonumber \\
  g_{D D_s K^*} &=& \sqrt2 g_{PPV}, \nonumber \\
  g_{\phi D_s D_s} &=& -\sqrt2 g_{PPV}, \nonumber \\
  g_{J/\psi D_s D_s} &=& -\sqrt2 g_{PPV}, \nonumber \\
  g_{D_s^* K D^*} &=& g_{VVP}, \nonumber \\
  g_{D_s^* D K^*} &=& g_{VVP}, \nonumber \\
  g_{D_s^* D_s J/\psi} &=& g_{VVP}, \nonumber \\
  g_{D_s^* D_s \phi} &=& - g_{VVP}, \nonumber \\
  g_{D_s^* \eta D_s^*} &=& - \sqrt{\frac23} g_{VVP}, \nonumber \\
  g_{D_s^* \eta^\prime D_s^*} &=& \frac1{\sqrt3} g_{VVP}, \nonumber \\
  g_{D_s^* \eta_c D_s^*} &=& g_{VVP}, \nonumber \\
  g_{D^* \pi D^*} &=& g_{VVP}, \nonumber \\
  g_{D^* \eta D^*} &=& \frac1{\sqrt6} g_{VVP}, \nonumber \\
  g_{D^* \eta^\prime D^*} &=& \frac1{\sqrt3} g_{VVP}, \nonumber \\
  g_{D^* \eta_c D^*} &=& g_{VVP}, \nonumber \\
  g_{D^* D J/\psi} &=& g_{VVP}, \nonumber \\
  g_{D^* D \omega} &=& \frac1{\sqrt2} g_{VVP}, \nonumber \\
  g_{D^* D \rho} &=& g_{VVP}, \nonumber \\
  g_{D^* D_s K^*} &=& g_{VVP}, \nonumber \\
  g_{D_s^* J/\psi D_s^*} &=& g_{VVV}, \nonumber \\
  g_{D_s^* \phi D_s^*} &=& g_{VVV}, \nonumber \\
  g_{D_s^* K^* D^*} &=& - g_{VVV}, \nonumber \\
  g_{D^* J/\psi D^*} &=& g_{VVV}, \nonumber \\
  g_{D^* \rho D^*} &=& - g_{VVV}, \nonumber \\
  g_{D^* \omega D^*} &=& - \frac1{\sqrt2} g_{VVV}, \nonumber \\
  g_{\Xi_c^* \Xi_c^\prime \pi} &=& \frac1{2\sqrt3} g_{DBP}, \nonumber \\
  g_{\Xi_c^* \Xi_c^\prime \eta} &=& -\frac1{6\sqrt2} g_{DBP}, \nonumber \\
  g_{\Xi_c^* \Xi_c^\prime \eta^\prime} &=& \frac13 g_{DBP}, \nonumber \\
  g_{\Xi_c^* \Xi_c \pi} &=& \frac12 g_{DBP}, \nonumber \\
  g_{\Xi_c^* \Xi_c \eta} &=& \frac{\sqrt3}{2\sqrt2} g_{DBP}, \nonumber \\
  g_{\Xi_c^* \Lambda_c K} &=& -\frac12 g_{DBP}, \nonumber \\
  g_{\Xi_c^* \Lambda D} &=& \frac12 g_{DBP}, \nonumber \\
  g_{\Xi_c^* \Sigma D} &=& \frac1{\sqrt6} g_{DBP}, \nonumber \\
  g_{\Xi_c^* \Xi D_s} &=& -\frac1{\sqrt6} g_{DBP}, \nonumber \\
  g_{\Xi_c^* \Xi_c^\prime \rho} &=& \frac1{2\sqrt3} g_{DBV}, \nonumber \\
  g_{\Xi_c^* \Xi_c^\prime \omega} &=& \frac1{2\sqrt6} g_{DBV}, \nonumber \\
  g_{\Xi_c^* \Lambda D^*} &=& \frac12 g_{DBV}, \nonumber \\
  g_{\Xi_c^* \Sigma D^*} &=& \frac1{\sqrt6} g_{DBV}, \nonumber \\
  g_{\Xi_c^* \Xi D_s^*} &=& -\frac1{\sqrt6} g_{DBV},
  \end{eqnarray}
  where $g_{BBP}=0.989$, $g_{BBV}=3.25$, $g_{PPV}=3.02$, $g_{VVP}=-7.07$, $g_{VVV}=2.298$, $g_{DBP}=2.127$ and $g_{DBV}=16.03$~\cite{Janssen:1996kx,Ronchen:2012eg}. Note that the $D$ in the left side of the equation represents the $D$ meson and the $D$ in the right side of the equation refers to a decuplet baryon.

\end{appendix}

\bibliographystyle{plain}

\end{document}